\documentclass[letterpaper]{article}
\usepackage{style}
\usepackage{times} 
\usepackage{helvet}
\usepackage{courier}  
\usepackage[hyphens]{url}
\usepackage{graphicx} 
\usepackage[round]{natbib}
\usepackage{balance}
\usepackage{caption}
\usepackage{algorithm}
\usepackage{algorithmic}
\usepackage{booktabs}
\usepackage{amsmath}
\usepackage{cleveref}
\usepackage{caption}
\usepackage{subcaption}

\title{A Heterogeneous Agent Model of Mortgage Servicing: An Income-based Relief Analysis}
\author{
    Deepeka Garg\textsuperscript{\rm 1}\thanks{equal contribution},
    Benjamin Patrick Evans\textsuperscript{\rm 1}\footnotemark[1],
    Leo Ardon\textsuperscript{\rm 1},
    Annapoorani Lakshmi Narayanan\textsuperscript{\rm 1},
    Jared Vann\textsuperscript{\rm 1},
    Udari Madhushani\textsuperscript{\rm 1},
    Makada Henry-Nickie\textsuperscript{\rm 2},
    Sumitra Ganesh\textsuperscript{\rm 1}
}
\affiliations {
    \textsuperscript{\rm 1} JP Morgan AI Research \\
    \textsuperscript{\rm 2} JPMorgan Chase Institute \\
}

\begin{document}

\maketitle

\begin{abstract}
Mortgages account for the largest portion of household debt in the United States, totaling around \$12 trillion nationwide. In times of financial hardship, alleviating mortgage burdens is essential for supporting affected households. The mortgage servicing industry plays a vital role in offering this assistance, yet there has been limited research modelling the complex relationship between households and servicers. To bridge this gap, we developed an agent-based model that explores household behavior and the effectiveness of relief measures during financial distress. Our model represents households as adaptive learning agents with realistic financial attributes. These households experience exogenous income shocks, which may influence their ability to make mortgage payments. Mortgage servicers provide relief options to these households, who then choose the most suitable relief based on their unique financial circumstances and individual preferences. We analyze the impact of various external shocks and the success of different mortgage relief strategies on specific borrower subgroups. Through this analysis, we show that our model can not only replicate real-world mortgage studies but also act as a tool for conducting a broad range of what-if scenario analyses. Our approach offers fine-grained insights that can inform the development of more effective and inclusive mortgage relief solutions.
\end{abstract}

\section{Introduction and Background}

Following the 2008 crisis and the reported failure of mortgage servicers to provide adequate assistance to borrowers \citep{mcnulty2019regulation}, an increased focus has been placed on the role of servicers \citep{levitin2011mortgage},
and departments such as the Consumer Financial Protection Bureau \citep{levitin2012consumer} (CFPB) have been established to oversee these processes. Additionally, since the Basel III regulations, the sale of Mortgage Servicing Rights (MSR) from the lender to a dedicated servicer is becoming more common, as it has become increasingly costly for lenders to hold the MSRs \citep{goodman2014oasis}. Understanding the role servicers play in the mortgage ecosystem, the effectiveness of different relief types, and the preferences of borrowers for these relief types, are essential for improving the assistance offered to households during times of financial distress, whether during recessions or individual distress.

The pioneering work of \cite{geanakoplos2012getting}, which successfully modeled the complexity, heterogeneity, and multi-agent nature of the US housing market with an agent-based model (ABM), led to the development of similar ABMs for the UK \citep{carro}, Spain \citep{carro2023taming}, Australia \citep{evans2021impact}, and Hungary \citep{merHo2023high}, among others \citep{axtell2022agent}. 

Despite servicers' growing importance in the modern mortgage ecosystem, their effect on borrowers' behavior and, more widely, on the dynamics of the housing market, the current ABM literature has not thoroughly investigated this component.
For example, the key focus of many of these works is in understanding lending policy \citep{geanakoplos2012getting, laliotis2020agent}, predicting pricing dynamics \citep{evans2021impact}, analyzing securitisation \citep{lauretta2018hidden}, or modelling contagion \citep{bookstaber2017agent}. 
Mortgage servicing remains under-explored. Therefore, in this work, we focus on the mortgage servicing section of the housing market, and the relationship this has with the households financial well-being, to provide insights into mortgage assistance and household behaviour during times of financial distress.

\section{Proposed Model}
\begin{figure*}[!htb]
    \captionsetup{justification=centering}
    \centering
    \includegraphics[width=.7\textwidth]{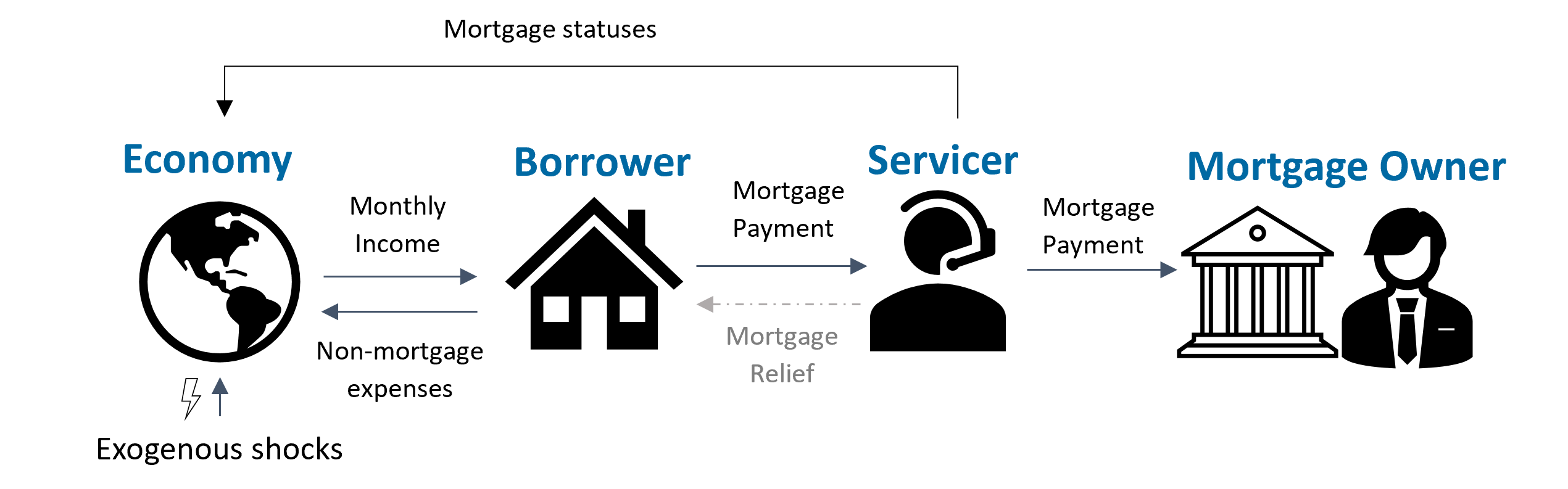}
    \caption{Agent interactions in the mortgage ecosystem.}
    \label{figProcess}
\end{figure*}

We develop an adaptive multi-agent model of the US servicing market using the \emph{Phantom} framework \citep{ardon2022phantom}. Each time step in the simulation represents one month, aligning with the typical mortgage payment schedule (one payment per month). We present in \Cref{figProcess} an overview of the agents modelled and how they interact in the simulation.

\subsection{Agents}

\subsubsection{Economy} The economy agent represents the outside ``economy". The economy is responsible for providing income to households, applying exogenous shocks, and updating house prices through a house price index $h$. During training, individual income shocks (applied to borrowers) arise from the economy on average once per year as a uniform random arrival process which may increase or reduce a borrower's income. During evaluation, we apply shocks of a particular size and analyze the impact on the ecosystem.

\begin{table}[]
\caption{Borrower financial characteristic data sources.}
\label{tblData}
\resizebox{\columnwidth}{!}{%
\begin{tabular}{lll}
\hline
\textbf{}                     & \textbf{Source} & \textbf{Conditioned on}        \\ \hline
\textbf{Income}               &      S1901 US Census           &                       \\
\textbf{Housing Expenses}     &   S2506 US Census              & Income \\
\textbf{Non-housing Expenses} &  Consumer Expenditure Survey               & Income \\
\textbf{Cash Savings}         &  Survey of Income and Program Participation               & Income \\ \hline
\end{tabular}
}
\end{table}

\subsubsection{Borrowers} Borrowers represent households holding mortgages at the start of the simulation\footnote{Borrowers can complete their mortgage during the simulation, either through foreclosure or paying off the loan.}. Borrowers receive income, pay housing and non-housing expenses, and accrue and spend savings. Borrowers have realistic heterogeneous financial characteristics sampled according to 2020 US census data and large-scale publicly available panel studies (\Cref{tblData}). Borrowers are strategic (learning via PPO \citep{schulman2017proximal}), making decisions to optimise their utility $U$ based on the market status and their financial characteristics. The borrowers' utility function is designed as a trade-off between a liquidity $L$ and equity $E$ component, parameterized by a liquidity preference $\gamma$:

\begin{equation}\label{eqReward}
    U = \gamma * L + (1-\gamma)*h*E
\end{equation}
where $L=1-\text{min}(1, \frac{\text{housing payment}}{\text{income}})$ and $E=\frac{\sum \text{loan payments made}}{\text{loan value}}$.

The equity component encodes desire for home-ownership, which has long been thought as the key factor in mortgage behaviour. However, recent studies \citep{farrell2019trading} have demonstrated the effect of liquidity on borrowers' decisions, motivating the inclusion of a liquidity component in the utility function. Borrowers having more available monthly liquidity allows for additional month-to-month consumption, expenditure, or savings. The liquidity preference (here, encoded by $\gamma$), is known to play an important role in strategic mortgage behaviour \citep{artavanis2023determinants}. $h$ represents the relative house price index, a numerical measure reflecting the current housing market conditions and varying the value of the equity owned by the borrowers.

\subsubsection{Servicer} The mortgage servicer manages all month-to-month loan activities, serving as the intermediary between the borrower and the mortgage owner. We model the servicer as following the Regulation X hierarchy\footnote{12 CFR Part 1024 - Real Estate Settlement Procedures Act}, based on procedures outlined by the CFPB. The servicer earns monthly fees, and during times of financial distress, is responsible for providing relief to borrowers, and advancing missed payments to mortgage owners, paying out of their own funds to do so. We detail the \textit{example} fees and costs used in this work in \Cref{tblServicerCosts}, where specifics may vary among servicers \citep{cordell2008incentives}. These costs are eventually recovered (up to some proportion based on $h$) throughout the loan or through property foreclosure.

\begin{table}[]
\caption{Servicing fees and costs.}\label{tblServicerCosts}
\resizebox{\columnwidth}{!}{%
\begin{tabular}{lll}
\hline
\multicolumn{1}{c}{}           & \textbf{Amount}            & \textbf{Source} \\ \hline
\textbf{Monthly servicing fee} & 0.025\% of monthly payment &      \cite{kaul2019options}           \\
\textbf{Advance payment cost}          &                            &      \cite{servicerFees}                \\
- General                      &   \$0                       &                 \\
- Distressed                   &    Missed payments (max $4$)                        &          \\
\textbf{Additional incentives}  &                            &  \cite{FannieMae}                \\
Repayment, Forbearance         & \$500                      &                \\
Loan Modification              & \$1000                     &                 \\
\textbf{Recovery proportion}   & min($h$, 1)                &                 \\ \hline
\end{tabular}
}
\end{table}

\subsubsection{Mortgage Owner} The mortgage owner is the generic entity owning the mortgage, generally a  financial institution (bank or non-bank) or a government-sponsored enterprise (GSE), such as Fannie Mae, Freddie Mac, etc. 

\hfill

\noindent 
Our mortgage servicing model is designed to be flexible, allowing for compartmentalised extensions, such as a different outside economy model (e.g., \citealt{hommes2022canvas}), or varying servicer behaviour based on specific mortgage owners.

\section{Experimental Results}
The three main metrics we analyse are: the proportion of borrowers missing at least one payment (the affected rate), the months before becoming affected, and the foreclosure rates, under different negative income shocks. One of the benefits of the proposed approach over representative agent approaches is the ability to break the analysis down into realistic subgroups of the population, so we not only analyse the overall rates, but also produce insights based on borrowers' characteristics. We analyse the impacts from both the borrower and the servicer's perspective. This analysis provides insights into what borrower segments are more susceptible to mortgage distress following an income shock, and how they benefit from the existing mortgage relief options. In doing so, we aim to improve relief across income bands and identify borrower segments that may require additional mortgage assistance.

\subsection{Borrowers}
\begin{figure}[!htb]
    \captionsetup{justification=centering}
    \centering
    \begin{subfigure}[b]{0.47\columnwidth}
\includegraphics[width=\textwidth]{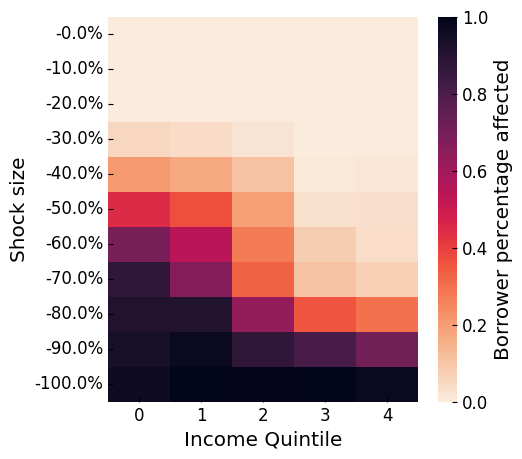}
    \caption{Affected rate}
    \label{figAffected}
    \end{subfigure}
    \hfill
    \begin{subfigure}[b]{0.47\columnwidth}
\includegraphics[width=.87\textwidth]{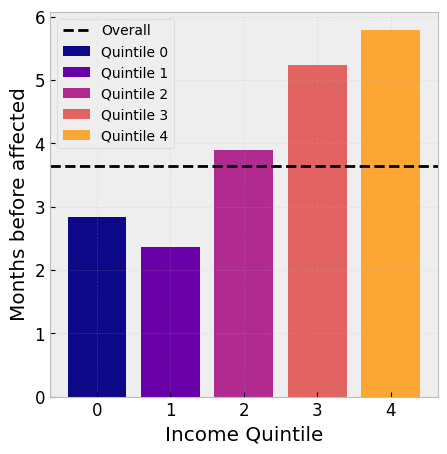}
    \caption{Average time-to-affect}
    \label{figAffectedTime}
    \end{subfigure}
    \hfill
    \caption{Impact of exogenous shocks on borrowers}
\end{figure}

The proportion of borrowers affected depends on the shock size and income quintile of the borrowers, \Cref{figAffected}. Despite applying the same \textit{relative} shock (e.g. a 20\% income reduction), lower-income borrowers are disproportionately affected, due to mortgage payments consuming a higher proportion of their income (leaving lower monthly liquidity). Additionally, this segment of borrowers generally have lower liquid cash savings, reducing the time-to-affect following an income shock (\Cref{figAffectedTime}). These results align with the real-world findings reported by \cite{farrell2018falling} based on extensive historical data analysis, both in terms of overall affect rates and time-to-affect, giving credence to the simulation, reproducing actual key trends through the simulated mortgage environment.

\subsection{Servicer}

\begin{figure}[!htb]
    \captionsetup{justification=centering}
    \centering
    \includegraphics[width=\columnwidth]{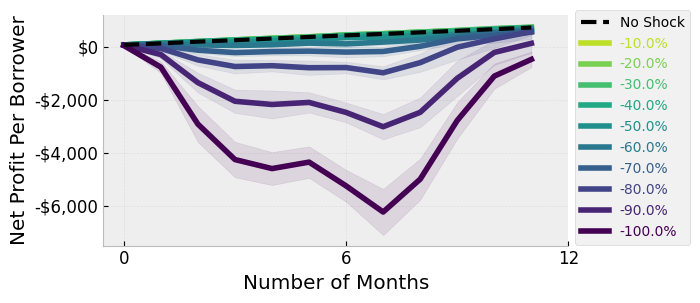}
    \caption{Net profit for the servicer (per borrower) under different shock sizes}
    \label{figExpectedProfit}
\end{figure}

From the servicer perspective, we track the net profit (based on \Cref{tblServicerCosts}) with and without income shocks (\Cref{figExpectedProfit}). Without shocks, the net profit is positive. Following larger shocks, servicers face temporary liquidity pressure (negative profit) due to the requirement of advancing missed mortgage payments to the mortgage owner. While these advanced funds are eventually recovered, this pressure poses a liquidity risk, and therefore is a vital consideration \citep{kim2018liquidity}, particularly when shocks are system-wide (e.g., recessions) \citep{davison2019temporary, kaul2020need}. 

Most of the servicers' profit comes from higher income quintiles, due to the servicing fee structure (based on a $0.25\%$ of the monthly payment, \Cref{tblServicerCosts}) and robustness of the higher income borrowers to the income shocks (\Cref{figAffected}). This analysis helps to confirm findings that servicers' focus on serving high-paying mortgages, to protect their cash flows \citep{diop2022mortgage}, and shows the importance of considering borrower heterogeneity: lower-income earners frequently face difficulties in acquiring and maintaining mortgages.  To foster financial inclusion, homeownership, and wealth building, developing products and policies that enhance the stability of lower-income households is essential \citep{advancingJP}. 

\section{Addons: Products}

The ABM serves as a scenario generator, allowing for counterfactual analysis. For example, with the proposed model, the impact of new products on borrowers financial wellbeing can be analysed. To this end, we consider the impact of one mortgage relief product, mortgage reserve accounts (MRA) \citep{goodman2023using}.

\subsection{Mortgage Reserve Accounts}

\begin{figure}[!htb]
     \captionsetup{justification=centering}
     \centering
     \begin{subfigure}[b]{0.47\columnwidth}
         \centering
         \includegraphics[width=\textwidth]{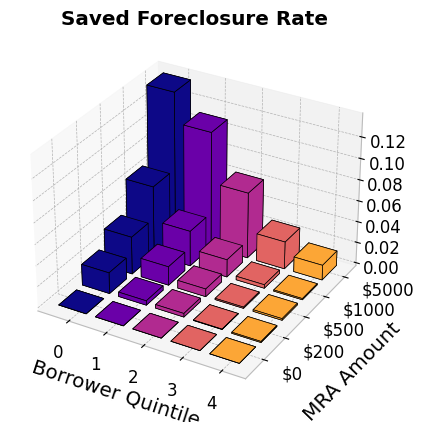}
         \caption{Foreclosure rates}
         \label{figMRAForeclose}
     \end{subfigure}
     \hfill
     \begin{subfigure}[b]{0.47\columnwidth}
         \centering
         \includegraphics[width=\textwidth]{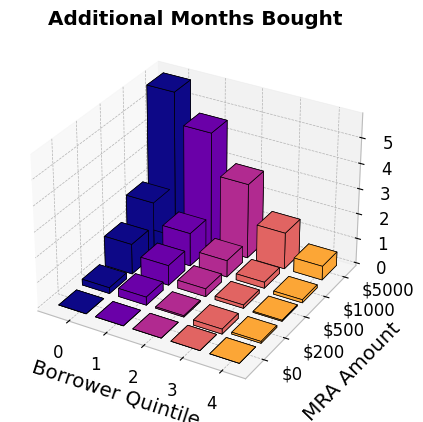}
         \caption{Time-to-affect}
         \label{figMRATime}
     \end{subfigure}
    \hfill
    \begin{subfigure}[b]{0.5\columnwidth}
             \centering
        \includegraphics[width=\textwidth]{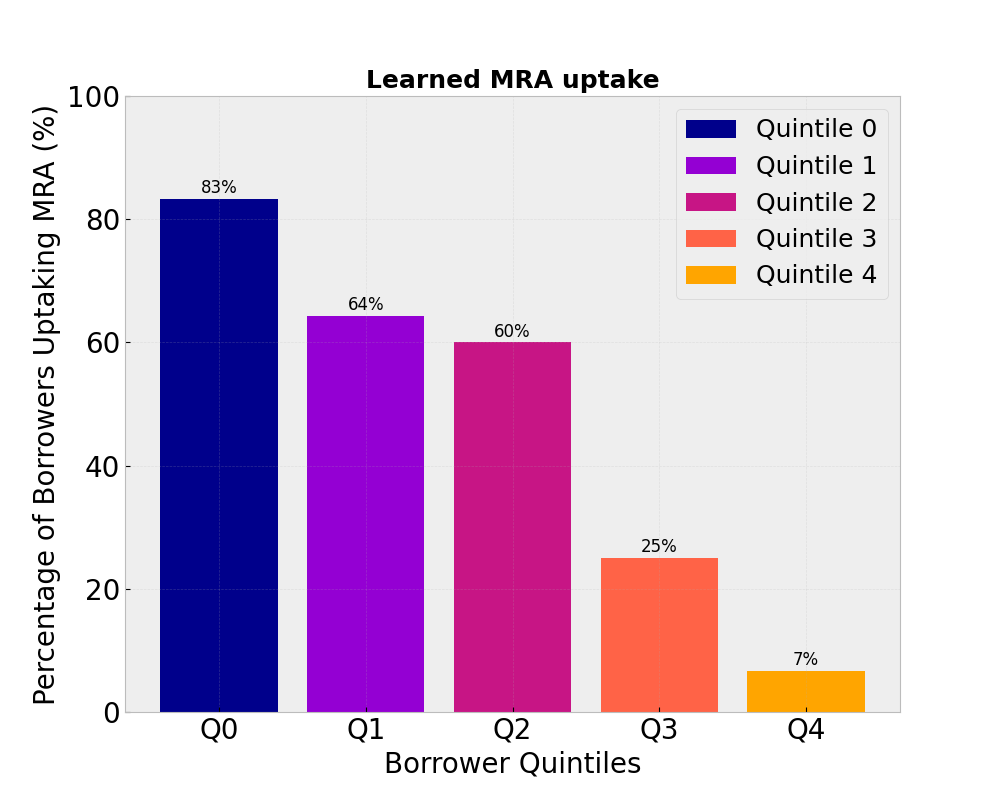}
    \caption{Matched MRA Uptake}
    \label{figMatchedMRA}
    \end{subfigure}
    \hfill
        \caption{Mortgage Reserve Accounts}
        \label{figMRA}
\end{figure}

We analyze a one-time fund of $\$M$ to cover missed mortgage payments. The saved foreclosures and additional months bought across $\$M$ are visualized in \Cref{figMRAForeclose,figMRATime}. Lower-income households find much greater assistance from this product, saving $12$ percentage points of foreclosures and providing up to 5 extra months with $M \to \$5000$.

\subsubsection{Matched MRA}

Rather than providing $\$M$ upfront, to incentivize borrower savings and prevent misuse, certain MRAs are provided on a savings match basis \citep{prosperityNow}, where borrowers put $\$m$ aside that is matched by the servicer, giving $\$M=2 \times \$m$.

In the simulation, borrowers learn whether to contribute to a matched MRA based on their expected utility (\Cref{eqReward}). The highest MRA uptake is seen among lower income borrowers  (\Cref{figMatchedMRA}), as they are more likely to face mortgage difficulties (\Cref{figAffected}) and therefore see more benefit from the product. These findings correspond with the results obtained from the MRA pilot study conducted with real households in the US \citep{prosperityNow}, showing lower income households agreed to partake and found use in the matched MRA.

\section{Conclusion}

We have developed a novel multi-agent model for mortgage servicing, addressing a crucial gap in existing housing market ABMs. This model enabled us to analyze the impact of exogenous income shocks on borrowers' capacity to meet their mortgage obligations, providing income-specific insights on timing, affect rates, and mortgage relief effectiveness. 
Our findings revealed that lower-income borrowers are disproportionately affected and are often unable to withstand more than a few months of reduced income. 
We validated the proposed model against actual studies conducted with human participants, successfully replicating important trends observed in the real world. 
Our approach's adaptability and the granularity of the resulting insights can be used to assess the impact of new mortgage products and contribute to the development of more effective and inclusive relief solutions through data-driven simulation.

\balance

\section*{Disclaimer}
This paper was prepared for informational purposes by the Artificial Intelligence Research group of JPMorgan Chase \& Co and its affiliates (“J.P. Morgan”) and is not a product of the Research Department of J.P. Morgan.  J.P. Morgan makes no representation and warranty whatsoever and disclaims all liability, for the completeness, accuracy or reliability of the information contained herein.  This document is not intended as investment research or investment advice, or a recommendation, offer or solicitation for the purchase or sale of any security, financial instrument, financial product or service, or to be used in any way for evaluating the merits of participating in any transaction, and shall not constitute a solicitation under any jurisdiction or to any person, if such solicitation under such jurisdiction or to such person would be unlawful. © 2024 JPMorgan Chase \& Co. All rights reserved.

\bibliographystyle{plainnat}
\bibliography{bib}

\begin{thebibliography}{30}
\providecommand{\natexlab}[1]{#1}
\providecommand{\url}[1]{\texttt{#1}}
\expandafter\ifx\csname urlstyle\endcsname\relax
  \providecommand{\doi}[1]{doi: #1}\else
  \providecommand{\doi}{doi: \begingroup \urlstyle{rm}\Url}\fi

\bibitem[Agava et~al.(2020)Agava, Davidof, and Ryan]{prosperityNow}
Pamela Agava, Kate Davidof, and Doug Ryan.
\newblock Mortgage reserve accounts: A mortgage-match savings initiative reducing default for low-income homeowners.
\newblock \emph{Prosperity Now White paper}, 2020.

\bibitem[Ardon et~al.(2023)Ardon, Vann, Garg, Spooner, and Ganesh]{ardon2022phantom}
Leo Ardon, Jared Vann, Deepeka Garg, Thomas Spooner, and Sumitra Ganesh.
\newblock Phantom-a rl-driven multi-agent framework to model complex systems.
\newblock In \emph{Proceedings of the 2023 International Conference on Autonomous Agents and Multiagent Systems}, pages 2742--2744, 2023.

\bibitem[Artavanis and Spyridopoulos(2023)]{artavanis2023determinants}
Nikolaos Artavanis and Ioannis Spyridopoulos.
\newblock Determinants of strategic behavior: Evidence from a foreclosure moratorium.
\newblock \emph{Journal of Financial Intermediation}, 56:\penalty0 101059, 2023.

\bibitem[Axtell and Farmer(2022)]{axtell2022agent}
Robert~L Axtell and J~Doyne Farmer.
\newblock Agent-based modeling in economics and finance: past, present, and future.
\newblock \emph{Journal of Economic Literature}, 2022.

\bibitem[Bookstaber(2017)]{bookstaber2017agent}
Richard Bookstaber.
\newblock Agent-based models for financial crises.
\newblock \emph{Annual Review of Financial Economics}, 9:\penalty0 85--100, 2017.

\bibitem[Carro(2023)]{carro2023taming}
Adrian Carro.
\newblock Taming the housing roller coaster: The impact of macroprudential policy on the house price cycle.
\newblock \emph{Journal of Economic Dynamics and Control}, page 104753, 2023.

\bibitem[Carro et~al.(2022)Carro, Hinterschweiger, Uluc, and Farmer]{carro}
Adrian Carro, Marc Hinterschweiger, Arzu Uluc, and J~Doyne Farmer.
\newblock {Heterogeneous effects and spillovers of macroprudential policy in an agent-based model of the UK housing market}.
\newblock \emph{Industrial and Corporate Change}, 32\penalty0 (2):\penalty0 386--432, 07 2022.
\newblock ISSN 0960-6491.
\newblock \doi{10.1093/icc/dtac030}.
\newblock URL \url{https://doi.org/10.1093/icc/dtac030}.

\bibitem[Cordell et~al.(2008)Cordell, Dynan, Lehnert, Liang, and Mauskopf]{cordell2008incentives}
Larry Cordell, Karen Dynan, Andreas Lehnert, Nellie Liang, and Eileen Mauskopf.
\newblock The incentives of mortgage servicers: Myths and realities.
\newblock 2008.

\bibitem[Davison(2019)]{davison2019temporary}
Lee Davison.
\newblock The temporary liquidity guarantee program: a systemwide systemic risk exception.
\newblock \emph{Journal of Financial Crises}, 1\penalty0 (2):\penalty0 1--39, 2019.

\bibitem[Diop and Zheng(2022)]{diop2022mortgage}
Moussa Diop and Chen Zheng.
\newblock Mortgage servicing fees and servicer incentives during loss mitigation.
\newblock \emph{Management Science}, 2022.

\bibitem[Evans et~al.(2021)Evans, Glavatskiy, Harr{\'e}, and Prokopenko]{evans2021impact}
Benjamin~Patrick Evans, Kirill Glavatskiy, Michael~S Harr{\'e}, and Mikhail Prokopenko.
\newblock The impact of social influence in australian real estate: Market forecasting with a spatial agent-based model.
\newblock \emph{Journal of Economic Interaction and Coordination}, pages 1--53, 2021.

\bibitem[{Fannie Mae}(2020)]{FannieMae}
{Fannie Mae}.
\newblock Lender letter (ll-2020-09), 2020.
\newblock URL \url{https://singlefamily.fanniemae.com/media/23091/display}.

\bibitem[Farrell et~al.(2018)Farrell, Bhagat, and Zhao]{farrell2018falling}
Diana Farrell, Kanav Bhagat, and Chen Zhao.
\newblock Falling behind: Bank data on the role of income and savings in mortgage default.
\newblock \emph{Available at SSRN 3273062}, 2018.

\bibitem[Farrell et~al.(2019)Farrell, Bhagat, and Zhao]{farrell2019trading}
Diana Farrell, Kanav Bhagat, and Chen Zhao.
\newblock Trading equity for liquidity: Bank data on the relationship between liquidity and mortgage default.
\newblock \emph{Farrell, Diana, Kanav Bhagat, and Chen Zhao}, 2019.

\bibitem[Geanakoplos et~al.(2012)Geanakoplos, Axtell, Farmer, Howitt, Conlee, Goldstein, Hendrey, Palmer, and Yang]{geanakoplos2012getting}
John Geanakoplos, Robert Axtell, Doyne~J Farmer, Peter Howitt, Benjamin Conlee, Jonathan Goldstein, Matthew Hendrey, Nathan~M Palmer, and Chun-Yi Yang.
\newblock Getting at systemic risk via an agent-based model of the housing market.
\newblock \emph{American Economic Review}, 102\penalty0 (3):\penalty0 53--58, 2012.

\bibitem[Getter(2022)]{servicerFees}
Darryl~E. Getter.
\newblock Mortgage servicing assets and selected market developments.
\newblock \emph{Congressional Research Service}, 2022.

\bibitem[Goodman and Lee(2014)]{goodman2014oasis}
Laurie Goodman and Pamela Lee.
\newblock Oasis: A securitization born from msr transfers.
\newblock \emph{Change}, 4:\penalty0 4Q13, 2014.

\bibitem[Goodman et~al.(2023)Goodman, Ratcliffe, Visalli, and Ballesteros]{goodman2023using}
Laurie Goodman, Janneke Ratcliffe, Katie Visalli, and Rita Ballesteros.
\newblock Using mortgage reserves to advance black homeownership.
\newblock \emph{Washington, DC: The Urban Institute}, 2023.

\bibitem[Hommes et~al.(2022)Hommes, He, Poledna, Siqueira, and Zhang]{hommes2022canvas}
Cars Hommes, Mario He, Sebastian Poledna, Melissa Siqueira, and Yang Zhang.
\newblock Canvas: A canadian behavioral agent-based model.
\newblock Technical report, Bank of Canada, 2022.

\bibitem[{JP Morgan Chase PolicyCentre}(2023)]{advancingJP}
{JP Morgan Chase PolicyCentre}.
\newblock Advancing affordable, sustainable homeownership.
\newblock \emph{The JPMorgan Chase PolicyCente}, 2023.

\bibitem[Kaul and Tozer(2020)]{kaul2020need}
Karan Kaul and Ted Tozer.
\newblock The need for a federal liquidity facility for government loan servicing.
\newblock \emph{Urban Institute}, 2020.

\bibitem[Kaul et~al.(2019)Kaul, Goodman, McCargo, and Hill-Jones]{kaul2019options}
Karan Kaul, Laurie Goodman, Alanna McCargo, and Todd Hill-Jones.
\newblock Options for reforming the mortgage servicing compensation model.
\newblock 2019.

\bibitem[Kim et~al.(2018)Kim, Laufer, Stanton, Wallace, and Pence]{kim2018liquidity}
You~Suk Kim, Steven~M Laufer, Richard Stanton, Nancy Wallace, and Karen Pence.
\newblock Liquidity crises in the mortgage market.
\newblock \emph{Brookings Papers on Economic Activity}, 2018\penalty0 (1):\penalty0 347--428, 2018.

\bibitem[Laliotis et~al.(2020)Laliotis, Buesa, Leber, and Poblaci{\'o}n]{laliotis2020agent}
Dimitrios Laliotis, Alejandro Buesa, Miha Leber, and Javier Poblaci{\'o}n.
\newblock An agent-based model for the assessment of ltv caps.
\newblock \emph{Quantitative Finance}, 20\penalty0 (10):\penalty0 1721--1748, 2020.

\bibitem[Lauretta(2018)]{lauretta2018hidden}
Eliana Lauretta.
\newblock The hidden soul of financial innovation: An agent-based modelling of home mortgage securitization and the finance-growth nexus.
\newblock \emph{Economic Modelling}, 68:\penalty0 51--73, 2018.

\bibitem[Levitin(2012)]{levitin2012consumer}
Adam~J Levitin.
\newblock The consumer financial protection bureau: An introduction.
\newblock \emph{Rev. Banking \& Fin. L.}, 32:\penalty0 321, 2012.

\bibitem[Levitin and Twomey(2011)]{levitin2011mortgage}
Adam~J Levitin and Tara Twomey.
\newblock Mortgage servicing.
\newblock \emph{Yale J. on Reg.}, 28:\penalty0 1, 2011.

\bibitem[McNulty et~al.(2019)McNulty, Garcia-Feijoo, and Viale]{mcnulty2019regulation}
James~E McNulty, Luis Garcia-Feijoo, and Ariel Viale.
\newblock The regulation of mortgage servicing: lessons from the financial crisis.
\newblock \emph{Contemporary Economic Policy}, 37\penalty0 (1):\penalty0 170--180, 2019.

\bibitem[M{\'e}r{\H{o}} et~al.(2023)M{\'e}r{\H{o}}, Borsos, Hossz{\'u}, Ol{\'a}h, and V{\'a}g{\'o}]{merHo2023high}
Bence M{\'e}r{\H{o}}, Andr{\'a}s Borsos, Zsuzsanna Hossz{\'u}, Zsolt Ol{\'a}h, and Nikolett V{\'a}g{\'o}.
\newblock A high-resolution, data-driven agent-based model of the housing market.
\newblock \emph{Journal of Economic Dynamics and Control}, 155:\penalty0 104738, 2023.

\bibitem[Schulman et~al.(2017)Schulman, Wolski, Dhariwal, Radford, and Klimov]{schulman2017proximal}
John Schulman, Filip Wolski, Prafulla Dhariwal, Alec Radford, and Oleg Klimov.
\newblock Proximal policy optimization algorithms.
\newblock \emph{arXiv preprint arXiv:1707.06347}, 2017.

\end{thebibliography}

\end{document}